\pdfoutput=1

\documentclass[11pt]{article}

\usepackage{misc/acl}
\usepackage{times}
\usepackage{latexsym}
\usepackage[T1]{fontenc}
\usepackage[utf8]{inputenc}
\usepackage[]{siunitx}
\usepackage{microtype}
\usepackage{amsmath}
\usepackage{amssymb}
\usepackage{bbm}
\usepackage{graphicx}
\usepackage{cleveref}
\usepackage{booktabs}
\usepackage{multirow}
\usepackage{enumitem}
\usepackage{multicol}
\usepackage{bm}
\usepackage{booktabs}

\usepackage{amsmath}
\DeclareMathOperator*{\argk}{arg\,top-k}
\DeclareMathOperator*{\argmax}{arg\,max}

\title{Knowledge Base Index Compression via Dimensionality \\ and Precision Reduction}
\author{
    Vilém Zouhar{\textnormal ,}\,\,
    Marius Mosbach{\textnormal ,}\,\,
    Miaoran Zhang \and
    Dietrich Klakow \\
    Department of Language Science \& Technology, Saarland Informatics Campus,\\
    Saarland University, Germany \\   
    \texttt{\{vzouhar,mmosbach,mzhang,dklakow\}@lsv.uni-saarland.de}
}

\begin{document}
\maketitle
\begin{abstract}

Recently neural network based approaches to knowledge-intensive NLP tasks, such as question answering, started to rely heavily on the combination of neural retrievers and readers.
Retrieval is typically performed over a large textual knowledge base (KB) which requires significant memory and compute resources, especially when scaled up.
On HotpotQA we systematically investigate reducing the size of the KB index by means of dimensionality (sparse random projections, PCA, autoencoders) and numerical precision reduction.

Our results show that PCA is an easy solution that requires very little data and is only slightly worse than autoencoders, which are less stable.
All methods are sensitive to pre- and post-processing and data should always be centered and normalized both before and after dimension reduction.
Finally, we show that it is possible to combine PCA with using 1bit per dimension.
Overall we achieve (1) 100$\times$ compression with 75\%, and (2) 24$\times$ compression with 92\% original retrieval performance.


\end{abstract}

\section{Introduction}


Recent approaches to knowledge-intensive NLP tasks combine neural network based models with a retrieval component that leverages dense vector representations \citep{guu2020realm,lewis2020retrieval,petroni2021kilt}.
The most straightforward example is question answering, where the retriever receives as input a question and returns relevant documents to be used by the reader (both encoder and decoder), which outputs the answer \citep{chen2020neural}.
The same approach can also be applied in other contexts, such as fact-checking \citep{tchechmedjiev2019claimskg} or knowledgable dialogue \citep{dinan2018wizard}.
Moreover, this paradigm can also be applied to systems that utilize e.g. caching of contexts from the training corpus to provide better output, such as the k-nearest neighbours language model proposed by \citet{khandelwal2019generalization} or the dynamic gating language model mechanism by \citet{yogatama2021adaptive}.
All these pipelines are generalized as retrieving an artefact from a knowledge base \citep{zouhar2021artefact} on which the reader is conditioned together with the query.

Crucially, all of the previous examples rely on the quality of the retrieval component and the knowledge base.
The knowledge base is usually indexed by dense vector representations\footnote{Sparse representations via BM25 \citep{robertson1995okapi} are also commonly used but not the focus of this work.} and the retrieval component performs maximum similarity search, commonly using the inner product or the $L^2$ distance, to retrieve documents\footnote{We refer to the retrieved objects as documents though they commonly range from spans of text (e.g. 100 tokens) to the full documents.} from the knowledge base.
Only the index alone takes up a large amount of size of the knowledge base, making deployment and experimentation very difficult.
The retrieval speed is also dependent on the dimensionality of the index vector.
An example of a large knowledge base is the work of \citet{borgeaud2021improving} which performs retrieval over a database of 1.8 billion documents.

This paper focuses on the issue of compressing the index through dimensionality and precision reduction and makes the following contributions:

\begin{itemize}[noitemsep]
    \item Comparison of various unsupervised index compression methods for retrieval, including random projections, PCA, autoencoder, precision reduction and their combination. 
    \item Examination of effective pre- and post-processing transformations, showing that centering and normalization are necessary for boosting the performance.
    \item Analysis on the impact of adding irrelevant documents and retrieval errors. Recommendations for use by practicioners.  
\end{itemize}


In \Cref{sec:methods}, we describe the problem scenario and the experimental setup. We discuss the results of different compression methods in \Cref{sec:model_comparison}. We provide further analysis in \Cref{sec:analysis} and conclude with usage recommendations in \Cref{sec:discussion}.
The repository for this project is available open-source.\footnote{
\href{https://github.com/zouharvi/kb-shrink}{github.com/zouharvi/kb-shrink}
}

\section{Related Work}

\paragraph{Reducing index size.}

A thorough overview of the issue of dimensionality reduction in information retrieval in the context of dual encoders has been done by \citet{luan2020sparse}.
Though in-depth and grounded in formal arguments, their study is focused on the limits and properties of dimension reduction in general (even with sparse representations) and the effect of document length on performance.
In contrast to their work, this paper aims to compare more methods and give practical advice with experimental evidence.

A baseline for dimensionality reduction has been recently proposed by \citet{izacard2020memory} in which they perform the reduction while training the document (and query) encoder by adding a low dimensional linear projection layer as the final output layer. Compared to our work, their approach is supervised.

In the concurrent work of \citet{ma2021simple}, PCA is also used to reduce the size of the document index.
Compared to our work, they perform PCA using the combination of all question and document vectors. We show in \Cref{fig:pca_auto_main,fig:model_data} that this is not needed and the PCA transformation matrix can be estimated much more efficiently. Moreover, we use different unsupervised compression approaches for comparison and perform additional analysis of our findings.

An orthogonal approach to the issue of memory cost has been proposed by \citet{yamada2021efficient}.
Instead of moving to another continuous vector representation, their proposed method maps original vectors to vectors of binary values which are trained using the signal from the downstream task.
The pipeline, however, still relies on re-ranking using the uncompressed vectors.
This method is different from ours and in \Cref{subsec:prec_reduction} we show that they can be combined. 

Finally, \citet{he2021efficient} investigate filtering and k-means pruning for the task of kNN language modelling. This work also circumvents the issue of having to always perform an expensive retrieval of a large data store by determining whether the retrieval is actually needed for a given input.

\paragraph{Effect of normalization.}

\citet{timkey2021all} examine how dominating embedding dimensions can worsen retrieval performance.
They study the contribution of individual dimensions find that normalization is key for document retrieval based on dense vector representation when BERT-based embeddings are used.
Compared to our work, they study pre-trained BERT directly, while we focus on DPR.

\section{Setup} \label{sec:methods}

\subsection{Problem Statement and Evaluation}

Given a query $q$, the following set of equations summarizes the conceptual progression from retrieving top $k$ relevant documents $Z = \{d_1, d_2, \ldots, d_k\}$ from a large collection of documents $\mathcal{D}$ so that the relevance of $d$ with $q$ is maximized.
For this, the query and the document embedding functions $f_Q : \mathcal{Q} \rightarrow \mathbb{R}^d$ and $f_D : \mathcal{D} \rightarrow \mathbb{R}^d$ are used to map the query and \textit{all} documents to a shared embedding space and a similarity function $\text{sim} : \mathbb{R}^d \times \mathbb{R}^d \rightarrow \mathbb{R}$ approximates the relevance between query and documents.
Here, we consider either the inner product or the $L^2$ distance as $\text{sim}$.\footnote{Cosine similarity could also be used but for computation reasons we skip it.
Results are the same as for inner product and $L^2$ distance when the vectors are normalized.}
Finally, to speed up the similarity computation over a large set of documents and to decrease memory usage ($f_D$ is usually precomputed), we apply \textbf{dimension reduction functions} $r_Q : \mathbb{R}^d \rightarrow \mathbb{R}^{d'}$ and $r_D : \mathbb{R}^d \rightarrow \mathbb{R}^{d'}$ for the query and document embeddings respectively. Formally, we are solving the following problem:
\begin{align}
Z &= \argk_{d \in \mathcal{D}}~ \text{rel.}(q, d)~, \text{with}\\
\text{rel.}(q, d) &\approx \text{sim}(f_\text{Q}(q), f_\text{D}(d)) \\
&\approx \text{sim}(r_\text{Q}(f_\text{Q}(q)), r_\text{D}(f_\text{D}(d)))
\end{align}

The approximation in (2) was shown to work well in practice for inner product and $L^2$ distance \citep{lin2021proposed}.
In this case, $f_Q$ is commonly fine-tuned for a specific downstream task.
For this reason, it is desirable in (3) for the functions $r_Q$ and $r_D$ to be differentiable so that they can propagate the signal.
These dimension-reducing functions need not be the same because even though they project to a shared vector space, the input distribution may still be different.
Similarly to the query and document embedding functions, they can be fine-tuned.

\paragraph{Task Agnostic Representation.}

When dealing with multiple downstream tasks that share a single (large) knowledge base, typically only $f_Q$ is fine-tuned for a specific task while $f_D$ remains fixed \citep{lewis2020retrieval,petroni2021kilt}.
This assumes that the organization of the document vector space is sufficient across tasks and that only the mapping of the queries to this space needs to be trained.\footnote{\citet{guu2020realm} provide evidence that this assumption can lead to worse results in some cases.}
Hence, this work is motivated primarily by finding a good $r_D$ (because of the dominant size of the document index), though we note that $r_Q$ is equally important and necessary because even without any vector semantics, the key and the document embeddings must have the same dimensionality.



\paragraph{R-Precision.}

To evaluate retrieval performance we compute \textit{R-Precision} averaged over queries: (relevant documents among top $k$ passages in $Z$)$/r$, $k$ = number of passages in relevant documents, in the same way as \citet{petroni2021kilt}.
Following previous work, we consider the inner product (IP) and the $L^2$ distance as the similarity function.

\subsection{Data} \label{subsec:model_and_data}

As knowledge base we use documents from English Wikipedia and follow the setup described by \citet{petroni2021kilt}.
We mark spans (original articles split into 100 token pieces, 50 million in total) as relevant for a query if they come from the same Wikipedia article as one of the provenances.\footnote{Spans of the original text which help in answering the query.}
In order to make our experiments computationally feasible and easy to reproduce we experiment with a modified version of this knowledge base where we keep only spans of documents that are relevant to at least one query from the training or validation set of our downstream tasks. As downstream tasks, we use HotpotQA \citep{yang2018hotpotqa} for all main experiments and Natural Questions \citep{kwiatkowski2019natural} to verify that the results transfer to other datasets as well.
This leads to over 2 million encoded spans for HotpotQA (see \Cref{tab:dataset_size} for dataset sizes).
The 768-dimensional embeddings (32-bit floats) of this dataset (both queries and documents) add up to 7GB (146GB for the whole unpruned dataset).

\subsection{Uncompressed Retrieval Peformance}

To establish baselines for uncompressed performance we use models based on BERT \citep{devlin2019bert}.
We consider (1) vanilla BERT, (2) SentenceBERT \citep{reimers2019sentence} and (3) DPR \citep{karpukhin2020dense}, which was specifically trained for document retrieval.
To obtain document embeddings, we use either the last hidden state representation at \texttt{[CLS]} or the average across tokens of the last layer.

\begin{figure}[t]
    \includegraphics[width=\linewidth]{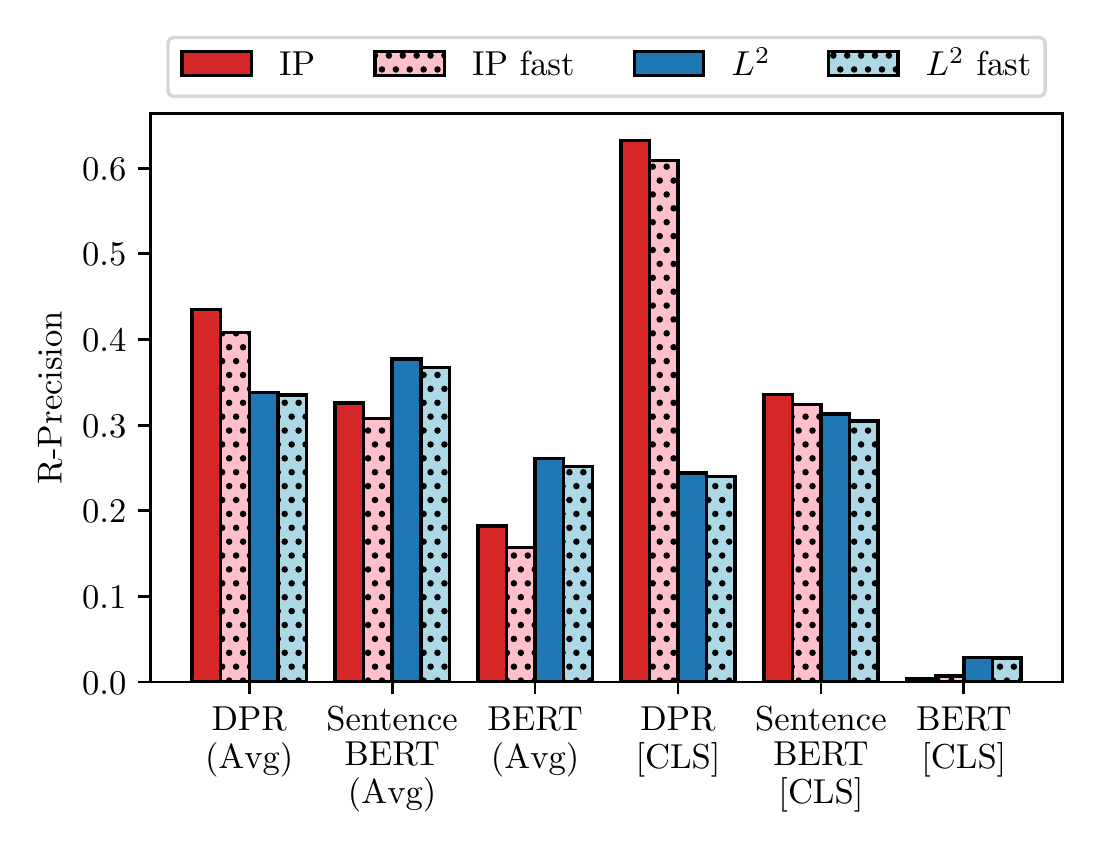}
    
    \caption{Comparison of different BERT-based embedding models and versions when using faster but slightly inaccurate nearest neighbour search. \textbf{[CLS]} is the specific token embedding from the last layer while \textbf{(Avg)} is all token average.}
    \label{fig:model_intro}
\end{figure}

Our first experiment compares the retrieval performance of the different models on HotpotQA. The result is shown in \Cref{fig:model_intro}.
In alignment with previous works \citep{reimers2019sentence} an immediately noticeable conclusion is that vanilla BERT has a poor performance, especially when taking the hidden state representation for the \texttt{[CLS]} token.
Next, to make computation tractable, we repeat the experiment using FAISS \citep{JDH17}.\footnote{{IndexIVFFlat, nlist=200, nprobe=100}.}
We find that the performance loss across models is systematic, which warrants the use of this approximate nearest neighbour search for comparisons and all our following experiments will use FAISS on the DPR-CLS model.

\begin{figure}[ht]
    \includegraphics[width=\linewidth]{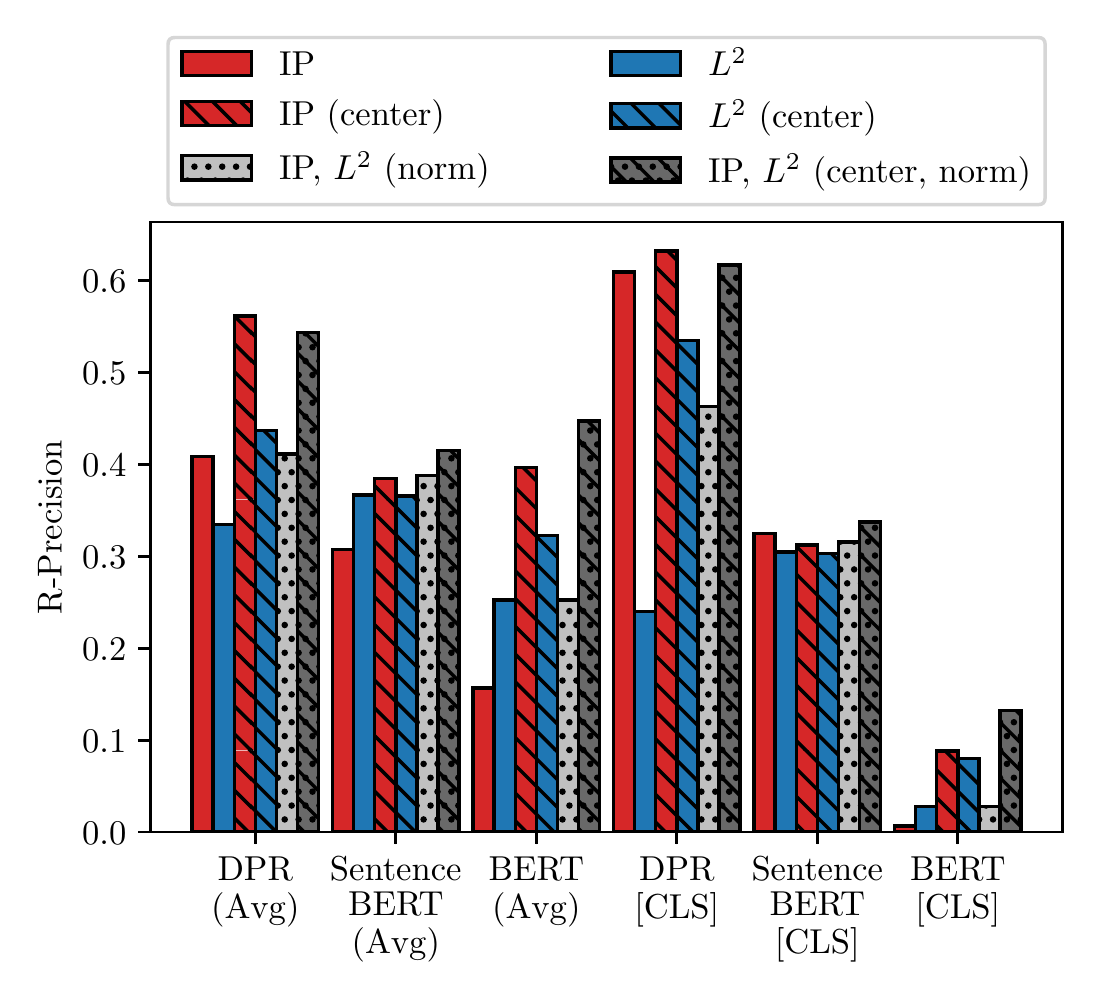}
    
    \caption{Effect of data centering and normalization on performance  (evaluated with FAISS).}
    \label{fig:model_normalization}
    
\end{figure}

\paragraph{Pre-processing Transformations.}

\Cref{fig:model_intro} also shows that model performance, especially for DPR, depends heavily on what similarity metric is used for retrieval.
This is because none of the models produces normalized vectors by default.

\Cref{fig:model_normalization} shows that performing only normalization ($\frac{\bm{x}}{||\bm{x}||}$) sometimes hurts the performance but when joined with centering beforehand ($\frac{\bm{x}-\bar{\bm{x}}}{||\bm{x}-\bar{\bm{x}}||}$), it improves the results (compared to no pre-processing) in all cases.
The normalization and centering is done for queries and documents separatedly.
Moreover, if the vectors are normalized, then the retrieved documents are the same for $L^2$ and inner product.
\footnote{$
\argmax_k -||\bm{a}-\bm{b}||^2 =
\argmax_k -\langle \bm{a},\bm{a} \rangle^2 - \langle \bm{b},\bm{b} \rangle^2 + 2\cdot \langle \bm{a},\bm{b} \rangle =
\argmax_k\ 2\cdot \langle \bm{a},\bm{b} \rangle - 2 = \argmax_k\ \langle \bm{a},\bm{b} \rangle$}

Nevertheless, we argue it still makes sense to study the compression capabilities of $L^2$ and the inner product separately, since the output of the compression of normalized vectors need not be normalized.


\section{Compression Methods} \label{sec:model_comparison}

Having established the retrieval performance of the uncompressed baseline, we now turn to methods for compressing the dense document index and the queries.

Note that we consider unsupervised methods on already trained index, for maximum ease of use and applicability.
This is in contrast to supervised methods, which have access to the query-doc relevancy mapping, or to in-training dimension reduction (i.e. lower final layer dimension).

\subsection{Random Projection}

The simplest way to perform dimension reduction for a given index $\bm{x} \in \mathbb{R}^d$ is to randomly preserve only certain $d'$ dimensions and drop all other dimensions:
\begin{gather*}
f_\text{drop.}(\bm{x}) = (x_{m_1}, x_{m_2}, \ldots, x_{m_{d'}})
\end{gather*}

Another approach is to greedily search which dimensions to drop (those that, when omitted, either improve the performance or lessen it the least):
\begin{align*}
& p_i(\bm{x}) = (x_0, x_1, {\ldots}, x_{i-1}, x_{i+1}, {\ldots}, x_{768}) \\
& \mathcal{L}_i = \text{R-Prec}(p_i(Q), p_i(D)) \\
& m = \text{sort}^\text{desc.}_\mathcal{L}([1 \ldots 768]) \\
& f_\text{greedy drop.}(\bm{x}) = (x_{m_1}, x_{m_2}, \ldots, x_{m_{d'}})
\end{align*}

The advantage of these two approaches is that they can be represented easily by a single $\mathbb{R}^{768\times d}$ matrix.
We consider two other standard random projection methods: Gaussian random projection and Sparse random projection \citep{fodor2002survey}.
Such random projections are suitable mostly for inner product \citep{kaski1998dimensionality} though the differences are removed by normalizing the vectors (which also improves the performance).

\begin{figure}[ht]
    \includegraphics[width=1\linewidth]{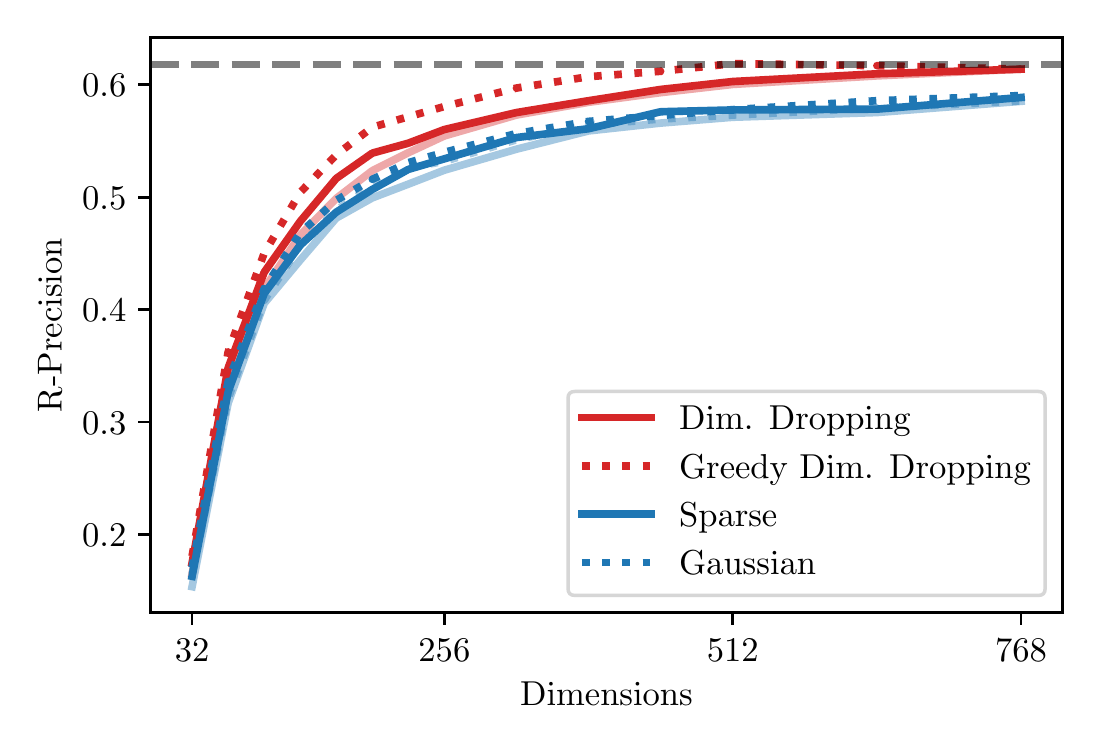}
    
    \caption{Dimension reduction using different random projections methods. Presented values are the max of 3 runs (except for greedy dimension dropping, which is deterministic), semi-transparent lines correspond to the minimum. Embeddings are provided by centered and normalized DPR-CLS. Final vectors are also post-processed by centering and normalization.}
    \label{fig:random_projection}
\end{figure}

\paragraph{Results.}

The results of all random projection methods are shown in \Cref{fig:random_projection}.
Gaussian random projection seems to perform equally to sparse random projection.
The performance is not fully recovered for the two methods.
Interestingly, simply dropping random dimensions led to better performance than that of sparse or Gaussian random projection.
The greedy dimension dropping even improves the performance slightly over random dimension dropping in some cases before saturating and is deterministic.
As shown in \Cref{tab:retrieval_summary}, the greedy dimension dropping with post-processing achieves the best performance among all random projection methods.
Without post-processing, $L_2$ distance works better compared to inner product. 

\begin{figure*}[ht]
    \centering
    \includegraphics[width=0.75\linewidth]{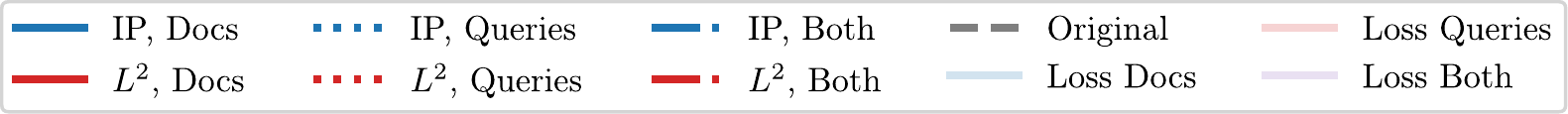}
    \includegraphics[width=\linewidth]{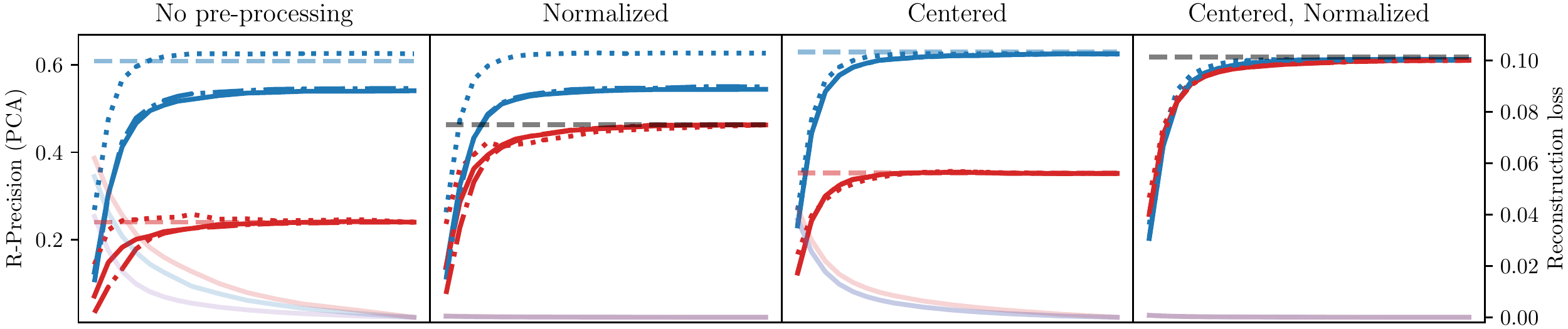}\\
    \includegraphics[width=\linewidth]{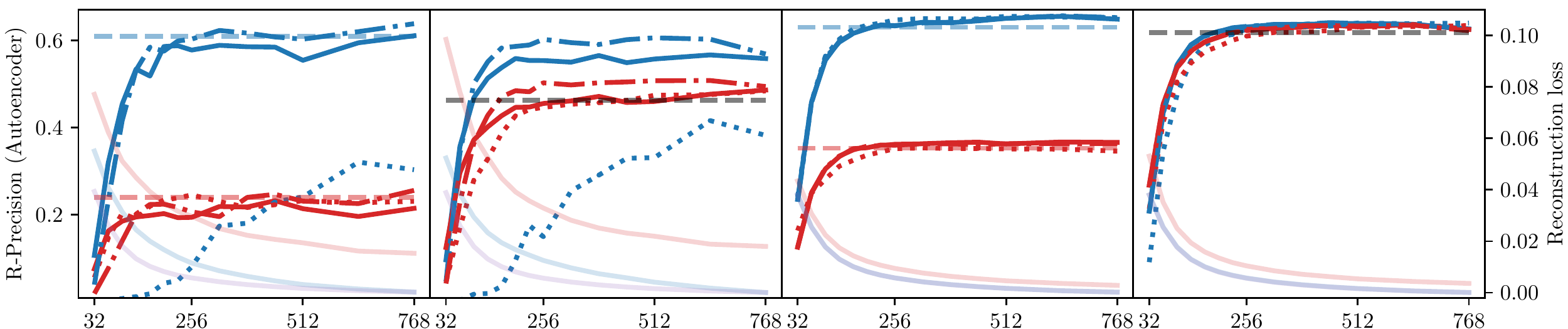}

    \caption{Dimension reduction using PCA (top) and Autoencoder (bottom) trained either on document index, query embeddings or both. Each figure corresponds to one of the four possible combinations of centering and normalizing the input data. The output vectors are not post-processed. Reconstruction loss (MSE, average for both documents and queries) is shown in transparent colour and computed in original data space. Horizontal lines show uncompressed performance. Embeddings are provided by DPR-CLS.}
    \label{fig:pca_auto_main}
\end{figure*}

\subsection{Principal Component Analysis}

Another natural candidate for dimensionality reduction is principal component analysis (PCA) \citep{pca}.
PCA considers the dimensions with the highest variance and omits the rest.
This leads to a projection matrix that projects the original data onto the principal components using an orthonormal basis $T$.
The following loss is minimized $\mathcal{L} = \text{MSE}(\mathbf{T'} \mathbf{T} \bm{x}, \bm{x})$.
Note that we fit PCA on the covariance matrix of either the document index, query embeddings or both and the trained dimension-reducing projection is then applied to both the document and query embeddings.

\paragraph{Results.} The results of performing PCA are shown in \Cref{fig:pca_auto_main}. First, we find that the uncompressed performance, as well as the effect of compression, is highly dependent on the data pre-processing.
This should not be surprising as the PCA algorithm assumes centered and pre-processed data. Nevertheless, we stress and demonstrate the importance of this step.
This is given by the normalization of the input vectors and also that the column vectors of PCA are orthonormal.

Second, when the data is not centered, the PCA is sensitive to what it is trained on.
\Cref{fig:pca_auto_main} show systematically that training on the set of available queries provides better performance than training on the documents or a combination of both. Subsequently, after centering the data, it does not matter anymore what is used for fitting: \textbf{both the queries and the documents provide good estimates of the data variance} and the dependency on training data size for PCA is explored explicitly in \Cref{subsec:model_comparison}.
The reason why queries provide better results without centering is that they are more centered in the first place, as shown in \Cref{tab:dataset_norm}.

\begin{table}[ht]
\centering
\resizebox{6.4cm}{!}{ %
\begin{tabular}{lcc}
 & \textbf{Avg.} $\mathbf{L^1}$ \textbf{(std)} & \textbf{Avg.} $\mathbf{L^2}$ \textbf{(std)} \\
\cmidrule{2-3}
\textbf{Documents} & $243.0\ (20.1)$ & $12.3\ (0.6)$ \\
\textbf{Queries} & $137.0\ (7.5)$ & $9.3\ (0.2)$
\end{tabular}
}
\caption{Average $L^1$ and $L^2$ norms of document and query embeddings from DPR-CLS without pre-processing.}
\label{tab:dataset_norm}
\end{table}

In all cases, the PCA performance starts to plateau around $128$ dimensions and is within $95\%$ of the uncompressed performance. 
Finally, we note that while PCA is concerned with minimizing reconstruction loss, \Cref{fig:pca_auto_main}  shows that even after vastly decreasing the reconstruction loss, no significant improvements in retrieval performance are achieved.
We further discuss this finding in \Cref{app_sec:reconstruction_loss}.

\paragraph{Component Scaling.}
One potential issue of PCA is that there may be dimensions that dominate the vector space.
\citet{mu2017all} suggest to simply remove the dimension corresponding to the highest eigenvalue though we find that simply scaling down the top k eigenvectors systematically outperforms standard PCA.
For simplicity, we focused on the top 5 eigenvectors and performed a small-scale grid-search of the scaling factors.
The best performing one was $(0.5, 0.8, 0.8, 0.9, 0.8)$ and \Cref{tab:retrieval_summary} shows that it provides a small additional boost in retrieval performance.

\subsection{Autoencoder} \label{subsec:auto}

A straightforward extension of PCA for dimensionality reducing is to use autoencoders, which has been widely explored \citep{hu2014improving,wang2016auto}.
Usually, the model is described by an encoder $e : \mathbb{R}^d \rightarrow \mathbb{R}^b$, a function from a higher dimension to the target (bottleneck) dimension and a decoder $r : \mathbb{R}^b \rightarrow \mathbb{R}^d$, which maps back from the target dimension to the original vector space.
The final (reconstruction) loss is then commonly computed as $\mathcal{L} = \text{MSE}((r \circ e)(\bm{x}), \bm{x})$.
To reduce the dimensionality of a dataset, only the function $e$ is applied to both the query and the document embedding.
We consider three models with the bottleneck:

\begin{enumerate}[itemsep=-1mm]
\item A linear projection similar to PCA but without the restriction of orthonormal columns:
\vspace{-1mm}
\begin{align*}
e_1(\bm{x}) &= L^{768}_{128} \\
r_1(\bm{x}) &= L^{128}_{768}
\end{align*}

\item A multi-layer feed forward neural network with $\tanh$ activation:
\vspace{-1mm}
\begin{align*}
e_2(\bm{x}) &= L^{768}_{512} \circ \tanh \circ L^{512}_{256} \circ \tanh \circ L^{256}_{128} \\
r_2(\bm{x}) &= L^{128}_{256} \circ \tanh \circ L^{256}_{512} \circ \tanh \circ L^{512}_{768}
\end{align*}

\item The same encoder as in the previous model but with a shallow decoder:
\vspace{-1mm}
\begin{align*}
e_3(\bm{x}) &= L^{768}_{512} \circ \tanh \circ L^{512}_{256} \circ \tanh \circ L^{256}_{128} \\ r_3(\bm{x}) &= L^{128}_{768}
\end{align*}
\end{enumerate}

Compared to PCA, it is able to model non-pairwise interaction between dimensions (in case of models 2 and 3 also non-linear interaction).

\paragraph{Results.}
We explore the effects of training data and pre-processing with results for the first model shown in \Cref{fig:pca_auto_main}.
Surprisingly, the Autoencoder is even more sensitive to proper pre-processing than PCA, most importantly centering which makes the results much more stable.

The rationale for the third model is that we would like the hidden representation to require as little post-processing as possible to become the original vector again.
The higher performance of the model with shallow decoder, shown in \Cref{tab:retrieval_summary} supports this reasoning.
An alternative way to reduce the computation (modelling dimension relationships) in the decoder is to regularize the weights in the decoder.
We make use of $L_1$ regularization explicitly because $L_2$ regularization is conceptually already present in Adam's weight decay.
This improves each of the three models.

Similarly to the other reconstruction loss-based method (PCA), without post-processing, inner product works yields better results.


\begin{table*}[ht]
\center
\resizebox{12.1cm}{!}{%
\begin{tabular}{lccccc}
\toprule
\multirow{2}{*}{\textbf{Method}} & \multirow{2}{*}{\textbf{Compression}} & \multicolumn{2}{c}{\textbf{Original}} & \textbf{Center + Norm.} \\ \cmidrule{3-5}
 &  & IP & $L^2$ & \{IP, $L^2$\} (\% original) \\
\midrule
Original & 1$\times$ & $0.609$ & $0.240$ & ${0.618}\,{\scriptstyle (100\%)}$ \\
\cmidrule{1-5}
Gaussian Projection (128) & 6$\times$ & $0.413$ & $0.453$ & $0.468\,{\scriptstyle (76\%)}$ \\
Sparse Projection (128) & 6$\times$ & $0.398$ & $0.448$& $0.457\,{\scriptstyle (74\%)}$ \\
Dimension Dropping (128) & 6$\times$ & $0.426$ & $0.466$ & $0.478\,{\scriptstyle (77\%)}$  \\
Greedy Dimension Dropping (128) & 6$\times$ & $0.447$ & $0.478$ & ${0.504}\,{\scriptstyle (82\%)}$  \\
\cmidrule{1-5}
PCA (128) & 6$\times$ & $0.577$ & $0.562$ & $0.579\,{\scriptstyle (94\%)}$ \\
PCA (128, scaled top 5) & 6$\times$ & $0.586$ & $0.572$ & ${0.592}\,{\scriptstyle (96\%)}$ \\ 
\cmidrule{1-5}
Autoencoder (128, single layer) & 6$\times$ & $0.585$ & $0.569$ & $0.588\,{\scriptstyle (95\%)}$ \\ 
Autoencoder (128, full) & 6$\times$ & $0.564$ & $0.560$ & $0.588\,{\scriptstyle (95\%)}$ \\ 
Autoencoder (128, shallow decoder) & 6$\times$ & $0.599$ & $0.582$ & $0.599\,{\scriptstyle (97\%)}$ \\ 
Autoencoder (128, single layer) + $L_1$ & 6$\times$ & $0.600$ & $0.587$ & ${0.601}\,{\scriptstyle (97\%)}$ \\ 
Autoencoder (128, full) + $L_1$ & 6$\times$ & $0.573$ & $0.569$ & $0.589\,{\scriptstyle (95\%)}$ \\ 
Autoencoder (128, shallow decoder) + $L_1$& 6$\times$ & $0.601$ & $0.591$ & ${0.601}\,{\scriptstyle (97\%)}$ \\ 
\cmidrule{1-5}
Precision 16-bit & 2$\times$ & $0.612$ & $0.610$ & $0.615\,{\scriptstyle (100\%)}$ \\
Precision 8-bit & 4$\times$ & $0.613$ & $0.610$ & $0.614\,{\scriptstyle (99\%)}$ \\
Precision 1-bit (offset $0.5$) & 32$\times$ & $0.559$ & $0.556$ & $0.561\,{\scriptstyle (91\%)}$ \\
Precision 1-bit (offset $0$) & 32$\times$ & $0.530$ & $0.556$ & $0.561\,{\scriptstyle (91\%)}$ \\
\cmidrule{1-5}
PCA (245) + Precision 1-bit (offset $0.5$)  & 100$\times$ & $0.459$ & $0.458$ & ${0.461}\,{\scriptstyle (75\%)}$ \\
PCA (128) + Precision 8-bit & 24$\times$ & $0.558$ & $0.553$ & ${0.567}\,{\scriptstyle (92\%)}$ \\
\bottomrule

\end{tabular}
}
\vspace{-0.15cm}
\caption{Overview of compression method performance (from 768) using either $L^2$ or inner product for retrieval. Inputs are based on centered and normalized output of DPR-CLS and the outputs optionally post-processed again. Performance is measured by R-Precision on HotpotQA.}
\vspace{-0.2cm}
\label{tab:retrieval_summary}
\end{table*}

\subsection{Precision Reduction} \label{subsec:prec_reduction}

Lastly, we also experiment with reducing index size by lowering the float precision from 32 bits to 16 and 8 bits.
Note that despite their quite high retrieval performance, they only reduce the size by 2 and 4 respectively (as opposed to 6 by dimension reduction via PCA to 128 dimensions).
Another drawback is that retrieval time is not affected because the dimensionality remains the same.

Using only one bit per dimension is a special case of precision reduction suggested by \citet{yamada2021efficient}.
Because we use centered data, we can define the element-wise transformation function as:
\begin{gather*}
f_\alpha(x_i) = \begin{cases}
1 - \alpha & x_i \ge 0 \\
0 -\alpha & x_i < 0
\end{cases}
\end{gather*}
Bit \texttt{1} would then correspond to $1-\alpha$ and \texttt{0} to $0-\alpha$.
While \citet{yamada2021efficient} use values $1$ and $0$, we work with $0.5$ and $-0.5$ in order to be able to distinguish between certain cases when using IP-based similarity.\footnote{When using $0$ and $1$, the IP similarity of $0$ and $1$ is the same as $0$ and $0$ while for $-0.5$ and $0.5$ they are $-0.25$ and $0.25$ respectively.}
As shown in \Cref{tab:retrieval_summary}, this indeed yields a slight improvement.
When applying post-processing, however, the two approaches are equivalent. While this method achieves extreme 32x compression on the disk and retains most of the retrieval performance, the downside is that if one wishes to use standard retrieval pipelines, these variables would have to be converted to a supported, larger, data type.\footnote{The Tevatron toolkit \citep{gao2022tevatron} supports mixed precision training with 16-bit floats.}

\subsection{Combination of PCA and Precision Reduction} \label{app_sec:prec_pca_mult}

\begin{figure}[ht]
\includegraphics[width=\linewidth]{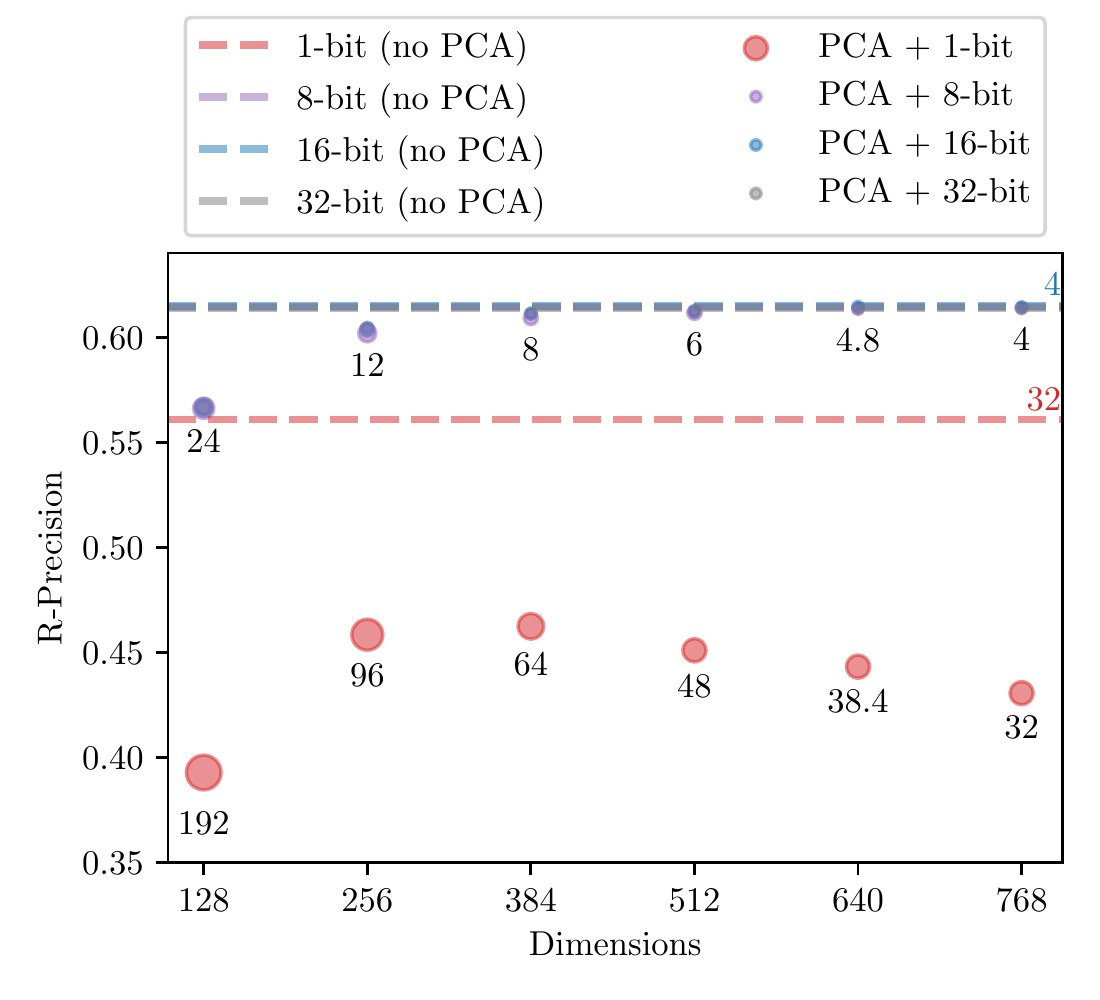}
\caption{Combination of PCA and precision reduction. Compression ratio is shown in text.
16-bit and 32-bit values overlap with 8-bit and their compression ratios are not shown. Measured on HotpotQA with DPR-CLS.}
\label{fig:prec_pca_mult}
\vspace{-0.26cm}
\end{figure}

Finally, reducing precision can be readily combined with dimension reduction methods, such as PCA (prior to changing the data type).
The results in \Cref{fig:prec_pca_mult} show that PCA can be combined with e.g. 8-bit precision reduction with negligible loss in performance.
As shown in the last row of \Cref{tab:retrieval_summary}, this can lead to the compressed size be 100x smaller while retaining $75\%$ retrieval performance on HotpotQA and $89\%$ for NaturalQuestions (see \Cref{tab:retrieval_summary_nq}).

\section{Analysis} \label{sec:analysis}

\subsection{Model Comparison} \label{subsec:model_comparison}

The comparison of all discussed dimension reduction methods is shown in \Cref{tab:retrieval_summary}.
It also shows the role of centering and normalization post-encoding which systematically improves the performance.
The best performing model for dimension reduction is the autoencoder with $L_1$ regularization and either just a single projection layer for the encoder and decoder or with the shallow decoder (6x compression with $97\%$ retrieval performance).
Additionally, \Cref{app_sec:speed} compares training and evaluation speeds of common implementations.

\subsection{Data size}

\begin{figure}[ht]
    \includegraphics[width=\linewidth]{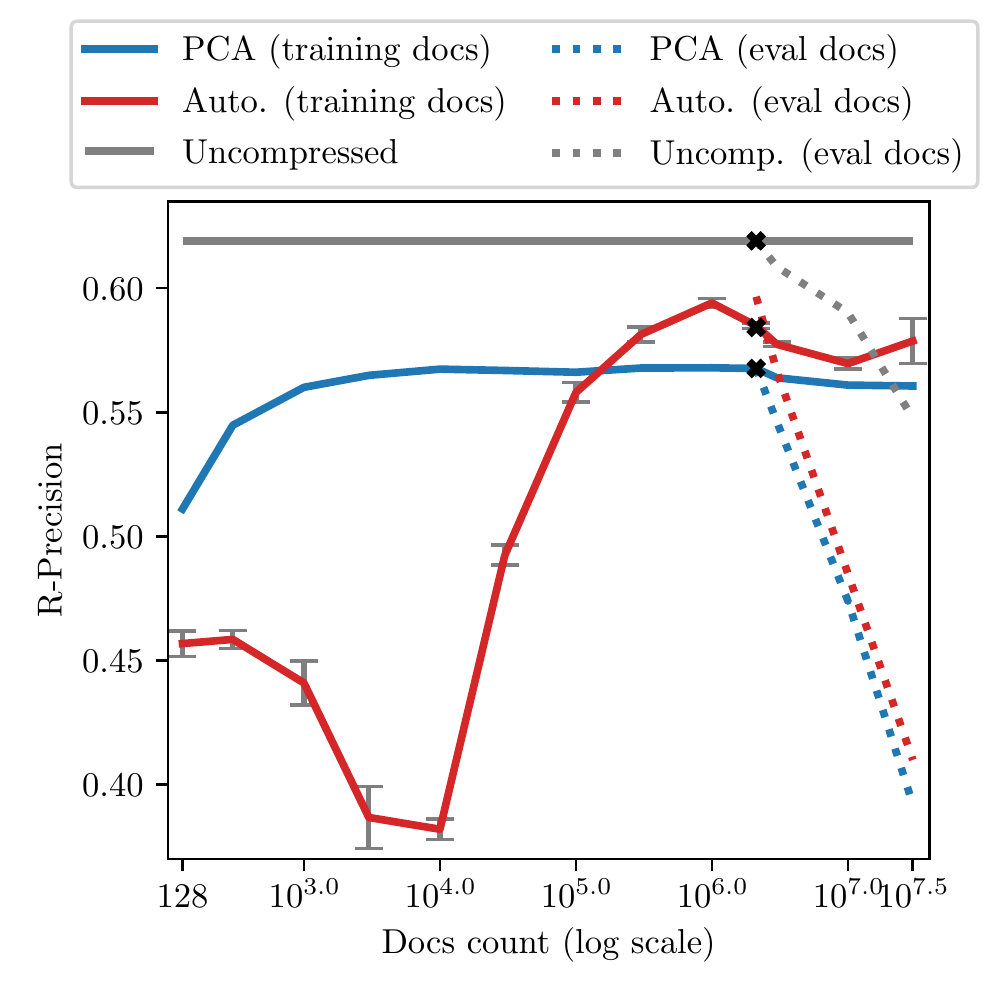}
    \caption{
    Dependency of PCA and autoencoder performance (evaluated on HotpotQA dev data, trained on document encodings, pre- and post-processing) by modifying the training data (solid lines) and by adding irrelevant documents to the retrieval pool (dashed lines). Black crosses indicate the original training size.
    Vertical bars are 95\% confidence intervals using t-distibution (across 6 runs with random model initialization and sampling).
    Note the log scale on the x-axis and the truncation of the y-axis.
    }
    \label{fig:model_data}
\end{figure}

A crucial aspect of the PCA and autoencoder methods is how much data they need for training.
In the following, we experimented with limiting the number of training samples for PCA and the linear autoencoder.
Results are shown in \Cref{fig:model_data}.

While \citet{ma2021simple} used a much larger training set to fit PCA, we find that PCA requires very few samples (lower-bounded by 128 which is also the number of dimensions used for this experiment).
This is because in the case of PCA training data is used to estimate the data covariance matrix which has been shown to work well when using a few samples \citep{tadjudin1999covariance}.
Additionally, we find that overall the autoencoder needs more data to outperform PCA.

Next, we experimented with adding more (potentially irrelevant) documents to the knowledge base. 
For this, we kept the training data for the autoencoder and PCA to the original size.
The results are shown as dashed lines in \Cref{fig:model_data}.
Retrieval performance quickly deteriorates for both models (faster than for the uncompressed case), highlighting the importance of filtering irrelevant documents from the knowledge base. 

\subsection{Retrieval errors}

So far, our evaluation focused on quantitative comparisons. In the following, we compare the distribution of documents retrieved before and after compression to investigate if there are systematic differences.
We carry out this analysis using HotpotQA which, by design, requires two documents in order to answer a given query.
We compare retrieval with the original document embeddings to retrieval with PCA and 1-bit compression.

\begin{figure}[ht]
\includegraphics[width=0.03\linewidth]{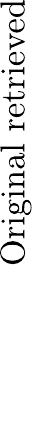}
\hspace{-0.2cm}
\includegraphics[width=0.502\linewidth]{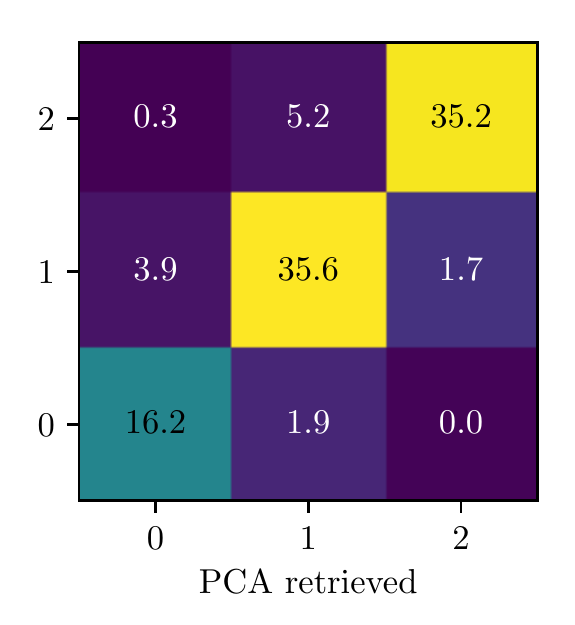}
\hspace{-0.39cm}
\includegraphics[width=0.502\linewidth]{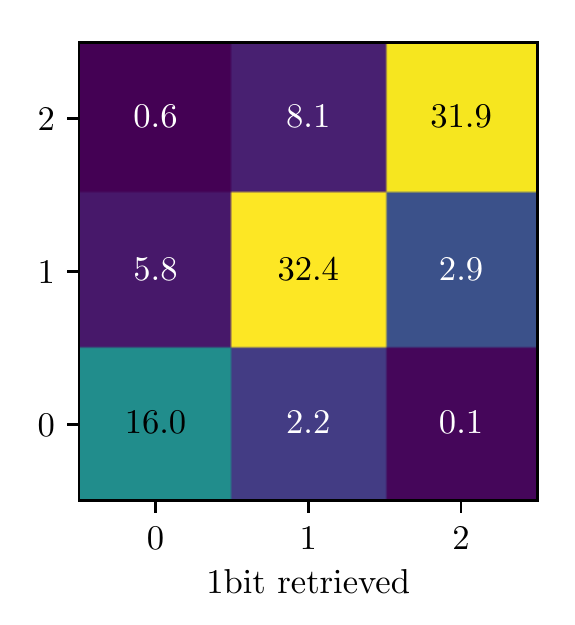}
\caption{Distribution of the number of retrieved documents for HotpotQA queries before and after compression: PCA (128) and 1-bit precision with R-Precisions (centered \& normalized) of $0.579$ and $0.561$, respectively.}
\label{fig:hits}
\end{figure}

We find that there are no systematic differences compared to the uncompressed retrieval.
This is demonstrated by the small off-diagonal values in \Cref{fig:hits}.
This result shows that if the retriever working with uncompressed embeddings returns two relevant documents in the top-k for a given query, also the retriever working with the compressed index is very likely to include the same two documents in the top-k.
This is further shown by the Pearson correlation in \Cref{tab:hits_correlation}.
This suggests that the compressed index can be used on downstream tasks with predictable performance loss based on the slightly worsened retrieval performance.
Furthermore, there do not seem to be any systematic differences even between the two vastly different compression methods used for this experiment (PCA and 1-bit precision).
This indicates that, despite their methodological differences, the two compression approaches seem to remove the same redundances in the uncompressed data. We leave a more detailed exploration of these findings for future work.

\subsection{Pitfalls of Reconstruction Loss} \label{app_sec:reconstruction_loss}

Despite PCA and autoencoder being the most successful methods, low reconstruction loss provides no theoretical guarantee to the retrieval performance.
Consider a simple linear projection that can be represented as a diagonal matrix that projects to a space of the same dimensionality.
This function has a trivial inverse and therefore no information is lost when it is applied.
The retrieval is however disrupted, as it will mostly depend on the first dimension and nothing else.
This is a major flaw of approaches that minimize the vector reconstruction loss because the optimized quantity is different to the actual goal.

\begin{center}
\resizebox{0.99\hsize}{!}{%
    $R = \begin{pmatrix}
    10^{99} & 0 & {\cdots} & 0 \\
    0 & 1 & {\cdots} & 0 \\
    {\vdots}  & {\vdots}  & {\ddots} & {\vdots}  \\
    0 & 0 & {\cdots} & 1
    \end{pmatrix}\quad R^{-1} = \begin{pmatrix}
    10^{-99} & 0 & {\cdots} & 0 \\
    0 & 1 & {\cdots} & 0 \\
    {\vdots}  & {\vdots}  & {\ddots} & {\vdots}  \\
    0 & 0 & {\cdots} & 1
    \end{pmatrix}$
}
\end{center}

\smallskip

\paragraph{Distance Learning.}
The task of dimensionality reduction has been explored by standard statistical methods by the name manifold learning.
The most used method is t-distributed stochastic neighbor (t-SNE) embedding built on the work of \citet{hinton2002stochastic} or multidimensional scaling \citep{kruskal1964nonmetric, borg2005modern}.
They organize a new vector space (of lower dimensionality) so that the $L^2$ distances follow those of the original space (extensions to other metrics also exist).
Although the optimization goal is more in line with our task of vector space compression with the preservation of nearest neighbours, methods of manifold learning are limited by the large computation costs\footnote{
The common fast implementation for t-SNE, Barnes-Hut \citep{barnes1986hierarchical, van2013barnes} is based on either quadtrees or octrees and is limited to $3$ dimensions.} and the fact that they do not construct a function but rather move the discrete points in the new space to lower the optimization loss.
This makes it not applicable for online purposes (i.e. adding new samples that need to be compressed as well).

The main disadvantage of the approaches based on reconstruction loss is that their optimization goal strays from what we are interested in, namely preserving distances between vectors.
We tried to reformulate the problem in terms of deep learning and gradient-based optimization to alleviate the issue of speed and extensibility of standard manifold learning approaches.
We try to learn a function that maps the original vector space to a lower-dimensional one while preserving similarities.
That can be either a simple linear projection $A$ or generally a more complex differentiable function $f$:
\begin{gather*}
\mathcal{L}=\text{MSE}(\text{sim}(f(t_i), f(t_j)), \text{sim}(t_i, t_j))
\end{gather*}

After the function $f$ is fitted, both the training and new data can be compressed by its application.
As opposed to manifold learning which usually leverages specific properties of the metrics, here they can be any differentiable functions.
The optimization was, however, too slow, underperforming (between sparse projection and PCA) and did not currently provide any benefits.

We also tried to use unsupervised contrastive learning by considering close neighbours in the original space as positive samples and distant neighbours as negative samples but reached similar results.

\section{Discussion} \label{sec:discussion}

In this section we briefly discuss the main conclusions from our experiments and analysis in the form of recommendations for NLP practicioners.

\paragraph{Importance of Pre-/post-processing.} As our results show, for all methods (and models), centering and normalization should be done before and after dimension reduction, as it boosts the performance of every model.

\paragraph{Method recommendation.} While most compression methods achieve similar retrieval performance and compression ratios (cf. \Cref{tab:retrieval_summary} and \Cref{tab:retrieval_summary_nq}), PCA stands out in the following regards: (1) It requires only minimal implementation effort and no tuning of hyper-parameters beyond selecting which principal components to keep; (2) as our analysis shows, the PCA matrix can be estimated well with only 1000 document or query embeddings. It is not necessary to learn a transformation matrix on the full knowledge base; (3) PCA can easily be combined with precision reduction based approaches.




\section{Summary} \label{sec:conclusion}

In this work, we examined several simple unsupervised methods for dimensionality reduction for retrieval-based NLP tasks: random projections, PCA, autoencoder and precision reduction and their combination.
We also documented the data requirements of each method and their reliance on pre- and post-processing.

\paragraph{Future work.}

As shown in prior works, dimension reduction can take place also during training where the loss is more in-line with the retrieval goal.
Methods for dimension reduction after training, however, rely mostly on reconstruction loss, which is suboptimal.
Therefore more research for dimension reduction methods is needed, such as fast manifold or distance-based learning.


\section*{Acknowledgements}

This work was funded by the Deutsche Forschungsgemeinschaft (DFG, German Research Foundation) – Project-ID 232722074 – SFB 1102.
Thank you to the reviewers, Badr M. Abdullah and many others for their comments to our work.

\bibliography{misc/bibliography}

\clearpage

\appendix

\begin{table}[ht]
\centering
\begin{tabular}{ll}
\toprule
\textbf{Hyperparameters} \\ \midrule
Batch size & 128 \\
Optimizer & Adam \\
Learning rate & 10$^{-3}$ \\
$L_1$ regularization & 10$^{-5.9}$ \\
\bottomrule
\end{tabular}
\caption{Hyperparameters of autoencoder architectures described in \Cref{subsec:auto}.
$L_1$ regularization is used only when explicitly mentioned.}
\label{app_sec:auto_config}
\end{table}

\begin{table}
    \center
    \begin{tabular}{lccc}
        \toprule
        & Uncompressed & PCA & 1bit \\
        \cmidrule{2-4}
        Uncompressed & $1.00$ &  &  \\
        PCA & $0.87$ & $1.00$ &  \\
        1bit & $0.81$ & $0.80$ & $1.00$ \\
        \bottomrule
    \end{tabular}
    \caption{Correlation of the number of retrieved documents for HotpotQA queries in different retrieval modes: uncompressed, PCA (128) and 1-bit precision with R-Precisions (centered \& normalized) of $0.618$, $0.579$ and $0.561$, respectively.}
    \label{tab:hits_correlation}
\end{table}

\section{Pre-processing} \label{app_sec:preprocessing}

Another common approach before any feature selection is to use z-scores ($\frac{x-\bar{x}}{\sigma}$) instead of the original values.
Its boost in performance is however similar to that of centering and normalization.
The effects of each pre-processing step are in \Cref{tab:zscore}.
The significant differences in performance show the importance of data pre-processing (agnostic to model selection).

\begin{table}[ht]
\centering
\resizebox{5cm}{!}{%
\begin{tabular}{lcccc}
    \toprule
     & \textbf{IP} & $\mathbf{L^2}$ \\
    \midrule
    DPR-CLS & $0.609$ & $0.240$ \\
    \cmidrule{1-3}
    Center & $0.630$ & $0.353$ \\
    Z-Score & $0.632$ & $0.525$ \\
    Norm. & \multicolumn{2}{c}{$0.463$} \\
    Center + norm. & \multicolumn{2}{c}{$0.618$} \\
    Z-Score + norm. & \multicolumn{2}{c}{$0.621$} \\
    \bottomrule
\end{tabular}
}
\caption{Effect of pre-processing transformations on embeddings produced by DPR-CLS. Means and standard deviations are computed separately for documents and queries. Transformation into z-scores includes centering.}
\label{tab:zscore}
\end{table}

\begin{figure*}[ht]
    \includegraphics[width=0.49\linewidth]{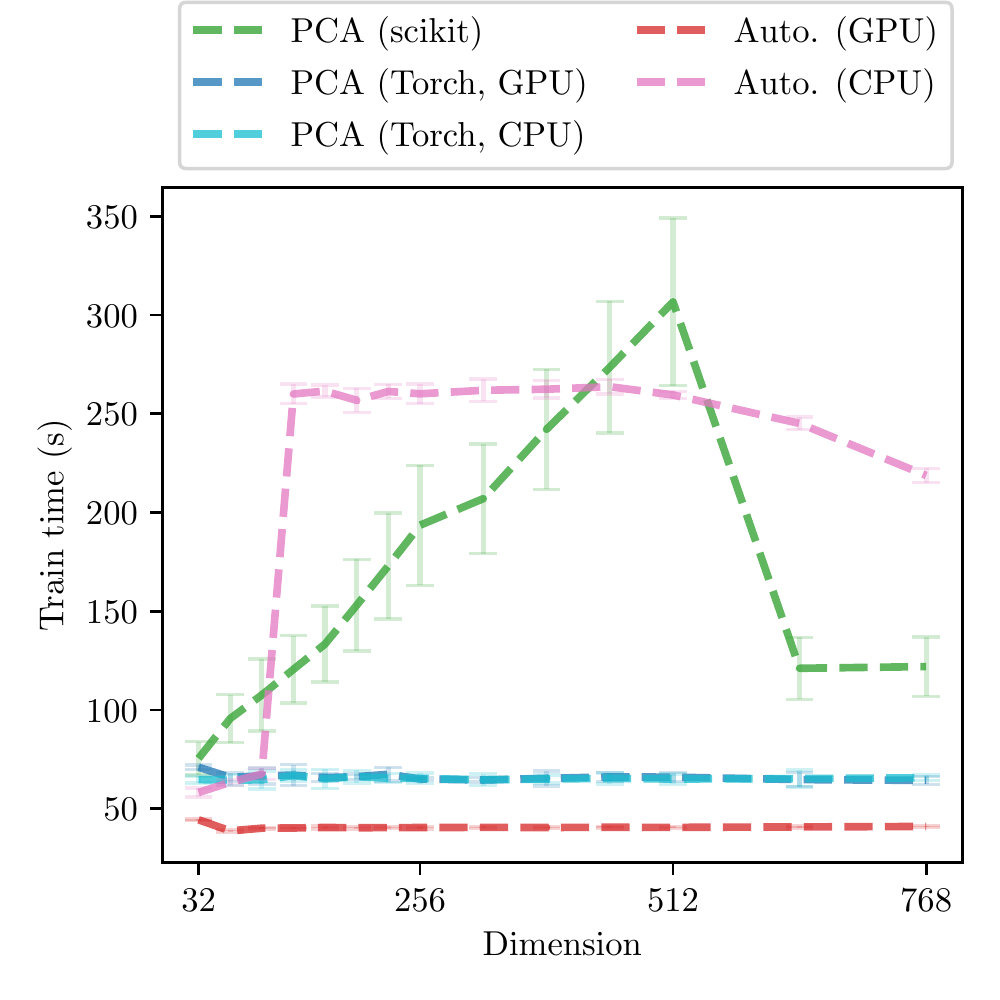}
    \includegraphics[width=0.49\linewidth]{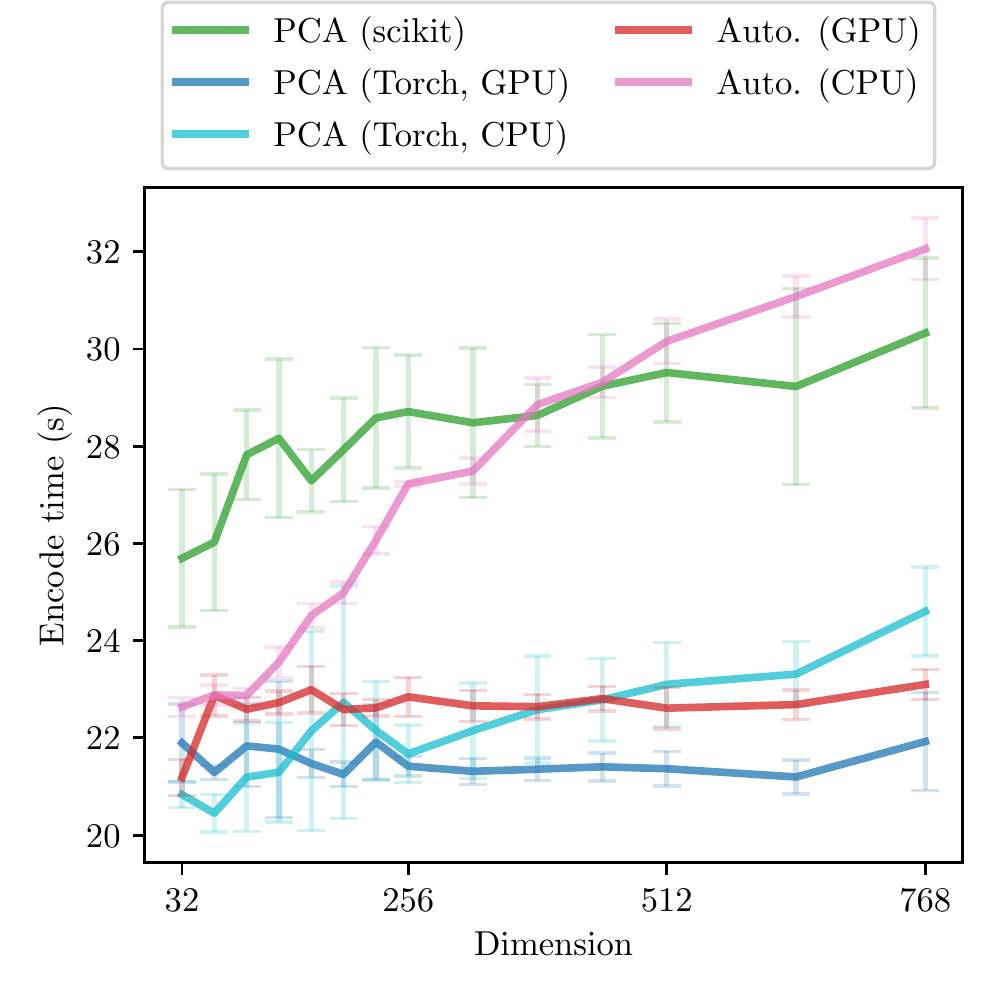}
    \caption{
    Speed comparison of PCA and autoencoder (model 3) implemented in PyTorch and Scikit\protect\footnotemark split into training and encoding parts. Models were trained on documents and queries jointly (normalized). Error bars are 95\% confidence intervals using t-distribution (5 runs).
    }
    \label{fig:model_speed}
\end{figure*}

\footnotetext{PyTorch 1.9.1, scikit-learn 0.23.2, RTX 2080 Ti (CUDA 11.4), 64$\times$2.1GHz Intel Xeon E5-2683 v4, 1TB RAM.}

\section{Speed} \label{app_sec:speed}

Despite the autoencoder providing slightly better retrieval performance and PCA being generally easier to use (due to the lack of hyperparameters), there are several tradeoffs in model selection.
Once the models are trained, the runtime performance (encoding) is comparable though for PCA it is a single matrix projection while for the autoencoder it may be several layers and activation functions.

Depending on the specific library used for implementation, however, the results differ.
\Cref{fig:model_speed} shows that the autoencoder (implemented in PyTorch) is much slower than any other model when run on a CPU but the fastest when run on a GPU.
Similarly, PCA works best if used from the PyTorch library (whether on CPU or GPU) and from the standard Scikit package.
Except for Scikit, there seems to be little relation between the target dimensionality and computation time.

\section{Comparison on Natural Questions} \label{app_sec:nq_model_comparison}

We also show the major experiments in \Cref{tab:retrieval_summary_nq} (table structure equivalent to that for the pruned dataset in \Cref{tab:retrieval_summary}) on Natural Question \citep{kwiatkowski2019natural} with identical dataset pre-processing.
The performance is overall larger because the task is different and the set of documents is lower (1.5 million spans) but comparatively the trends are in line with the previous conclusions of the paper.

\begin{table}
\centering
\begin{tabular}{lccc}
\toprule
Dataset & Train & Dev & Doc. \\
\midrule
HP & 69k & 6k & 49.7 Mio. \\
HP (pruned) & 69k & 6k & 2.1 Mio. \\
NQ (pruned) & 78k & 2k & 1.6 Mio. \\
\bottomrule
\end{tabular}
\caption{Number of training and dev queries and documents for HotpotQA and Natural Questions. Train and dev columns are queries.}
\label{tab:dataset_size}
\end{table}

\begin{table*}
\center
\resizebox{12.1cm}{!}{%
\begin{tabular}{lcccc}
\toprule
\multirow{2}{*}{\textbf{Method}} & \multirow{2}{*}{\textbf{Compression}} & \multicolumn{2}{c}{\textbf{Original}} & \textbf{Center + Norm.} \\ \cmidrule{3-5}
 &  & IP & $L^2$ & \{IP, $L^2$\} (\% original) \\
\midrule
Original & 1$\times$ & $0.934$ & $0.758$ & $0.920\,{\scriptstyle (100\%)}$ \\
\cmidrule{1-5}
Gaussian Projection & 6$\times$ & $0.825$ & $0.848$ & $0.855\,{\scriptstyle (93\%)}$ \\
Sparse Projection & 6$\times$ & $0.826$ & $0.848$& $0.856\,{\scriptstyle (93\%)}$ \\
Dimension Dropping & 6$\times$ & $0.840$ & $0.863$ & $0.867\,{\scriptstyle (94\%)}$ \\
Greedy Dimension Dropping & 6$\times$ & $0.845$ & $0.873$ & ${0.873}\,{\scriptstyle (95\%)}$  \\
\cmidrule{1-5}
PCA & 6$\times$ & $0.908$ & $0.907$ & $0.910\,{\scriptstyle (99\%)}$ \\
PCA (scaled top 5) & 6$\times$ & $0.916$ & $0.910$ & ${0.920}\,{\scriptstyle (100\%)}$ \\ 
\cmidrule{1-5}
Autoencoder (single layer) & 6$\times$ & $0.915$ & $0.910$ & $0.914\,{\scriptstyle (99\%)}$ \\ 
Autoencoder (full) & 6$\times$ & $0.903$ & $0.907$ & $0.910\,{\scriptstyle (99\%)}$ \\ 
Autoencoder (shallow decoder) & 6$\times$ & $0.916$ & $0.918$ & $0.919\,{\scriptstyle (100\%)}$ \\ 
Autoencoder + $L_1$ (single layer) & 6$\times$ & $0.918$ & $0.918$ & ${0.921}\,{\scriptstyle (100\%)}$ \\ 
Autoencoder + $L_1$ (full) & 6$\times$ & $0.909$ & $0.910$ & $0.913\,{\scriptstyle (99\%)}$ \\ 
Autoencoder + $L_1$ (shallow decoder) & 6$\times$ & $0.918$ & $0.917$ & ${0.919}\,{\scriptstyle (100\%)}$ \\ 
\cmidrule{1-5}
Precision 16-bit & 2$\times$ & $0.921$ & $0.917$ & $0.920\,{\scriptstyle (100\%)}$ \\
Precision 8-bit & 4$\times$ & $0.920$ & $0.921$ & $0.922\,{\scriptstyle (100\%)}$ \\
Precision 1-bit (offset $0.5$) & 32$\times$ & $0.902$ & $0.902$ & $0.904\,{\scriptstyle (98\%)}$ \\
Precision 1-bit (offset $0$) & 32$\times$ & $0.892$ & $0.902$ & $0.904\,{\scriptstyle (98\%)}$ \\
\cmidrule{1-5}
PCA (245) + Precision 1-bit (offset $0.5$)  & 100$\times$ & $0.854$ & $0.862$ & ${0.858}\,{\scriptstyle (93\%)}$ \\
PCA (128) + Precision 8-bit & 24$\times$ & $0.906$ & $0.904$ & ${0.909}\,{\scriptstyle (99\%)}$ \\
\bottomrule

\end{tabular}
}
\caption{Overview of compression method performance (from 768) using either $L^2$ or inner product for retrieval. Inputs are based on (1) original and (2) centered and normalized output of DPR-CLS. Performance is measured by R-Precision on NaturalQuestions.}
\label{tab:retrieval_summary_nq}
\end{table*}

\end{document}